\documentclass[11pt]{article}
\usepackage{graphicx}
\usepackage{amsmath}
\usepackage{amssymb}
\usepackage{amsxtra}
\def\Journal#1#2#3#4{{#1} {\bf #2} (#4) #3 }

\def\PLB{{ Phys. Lett.}  B}

\def\PRD{{ Phys. Rev.} D}

\def\GaC{ Gravitation and Cosmology}

\def\JETPL{ JETP Lett.}

\def\CQG{ Class. Quantum Grav.}

\def\IJMPA{{ Int. J. Mod. Phys.}  A}

\def\BWP{ Bled Workshops in Physics}

\def\MeV{\,{\rm MeV}}
\def\GeV{\,{\rm GeV}}
\def\TeV{\,{\rm TeV}}

\def\({\left(}
\def\){\right)}

\title{Search for Double Charged Particles as Direct Test for Dark Atom Constituents }
\author{O.~V.~Bulekov$^{1}$, M.~Yu.~Khlopov$^{1,2,3}$, A.~S.~Romaniouk$^{1,4}$, Yu.~S.~Smirnov$^{1}$\\$^{1}$National Research Nuclear University "MEPHI"\\(Moscow Engineering Physics Institute),\\ 115409 Moscow, Russia \\$^{2}$ Centre for Cosmoparticle Physics ``Cosmion"\\ 115409 Moscow, Russia \\$^{3}$ APC laboratory 10, rue Alice Domon et L\'eonie Duquet \\75205 Paris Cedex 13, France\\$^{4}$ CERN, Geneva, Switzerland}
\date{}

\begin{document}
\maketitle
\begin{center}

\end{center}

\begin{abstract}
The nonbaryonic dark matter of the Universe is assumed to consist of new stable particles.
Stable particle candidates for cosmological dark matter are usually considered as
neutral and weakly interacting. However stable charged leptons and quarks can also exist
hidden in elusive ”dark atoms” and can play a role of dark matter. Such possibility is
strongly restricted by the constraints on anomalous isotopes of light elements that form positively
charged heavy species with ordinary electrons. This problem might be avoided, if
stable particles with charge -2 exist and there are no stable particles with charges +1 and -1.
These conditions cannot be realized in supersymmetric models, but can be satisfied in several
 alternative scenarios, which are discussed in this paper. The excessive -2
charged particles are bound with primordial helium in O-helium ”atoms”, maintaining specific
nuclear-interacting form of the dark matter. O-helium dark matter can provide solution
for the puzzles of dark matter searches. The successful development of composite dark matter
scenarios appeals to experimental search for doubly charged constituents of dark atoms.
Estimates of production cross section of such particles at LHC are presented and discussed.
Signatures of double charged particles in the ATLAS experiment are outlined.
\end{abstract}

\section{Introduction}\label{intro}
The observation of exotic stable multi-charge objects would represent striking evidence for physics
beyond the Standard Model. Cosmological arguments put severe constraints on possible properties of
such objects. Such particles (see e.g. Ref. \cite{DMRev} for review and reference) should be stable, provide the measured dark matter density and be decoupled
from plasma and radiation at least before the beginning of matter dominated stage. The easiest way to
satisfy these conditions is to involve neutral elementary weakly interacting massive particles (WIMPs).
SUSY Models provide a list of possible WIMP candidates: neutralino, axino, gravitino etc.,
However it may not be the only particle physics solution for the dark matter problem.

One of such alternative solutions is based on the existence of heavy stable charged particles bound
in neutral ”dark atoms”. 
Dark atoms offer an interesting possibility to solve the puzzles of dark matter searches. It turns out that even stable electrically charged particles can exist hidden in such atoms,  bound by  ordinary Coulomb interactions (see \cite{DMRev,mpla,DDMRev} and references therein).
Stable particles with charge -1 are excluded due to overproduction of anomalous isotopes.  However, there doesn't appear such an evident contradiction for negatively doubly charged particles.

There
exist several types of particle models where heavy
stable -2  charged species, $O^{--}$, are predicted:
\begin{itemize}
\item[(a)] AC-leptons, predicted
as an extension of the Standard Model, based on the approach
of almost-commutative geometry \cite{Khlopov:2006dk,5,FKS,bookAC}.
\item[(b)] Technileptons and
anti-technibaryons in the framework of Walking Technicolor
(WTC) \cite{KK,Sannino:2004qp,Hong:2004td,Dietrich:2005jn,Dietrich:2005wk,Gudnason:2006ug,Gudnason:2006yj}.
\item[(c)] stable "heavy quark clusters" $\bar U \bar U \bar U$ formed by anti-$U$ quark of 4th generation
\cite{Khlopov:2006dk,Q,I,lom,KPS06,Belotsky:2008se} \item[(d)] and, finally, stable charged
clusters $\bar u_5 \bar u_5 \bar u_5$ of (anti)quarks $\bar u_5$ of
5th family can follow from the approach, unifying spins and charges\cite{Norma}.
\end{itemize}
All these models also
predict corresponding +2 charge particles. If these positively charged particles remain free in the early Universe,
they can recombine with ordinary electrons in anomalous helium, which is strongly constrained in
terrestrial matter. Therefore a cosmological scenario should provide a  mechanism which suppresses anomalous helium.
There are  two possible mechanisms than can provide a suppression:
\begin{itemize}
\item[(i)] The abundance of anomalous helium in the Galaxy may be significant, but in terrestrial matter
 a recombination mechanism could suppress this abundance below experimental upper limits \cite{Khlopov:2006dk,FKS}.
The existence of a new U(1) gauge symmetry, causing new Coulomb-like long range interactions between charged dark matter particles, is crucial for this mechanism. This leads inevitably to the existence of dark radiation in the form of hidden photons.
\item[(ii)] Free positively charged particles are already suppressed in the early Universe and the abundance
of anomalous helium in the Galaxy is negligible \cite{mpla,I}.
\end{itemize}
These two possibilities correspond to two different cosmological scenarios of dark atoms. The first one is
realized in the scenario with AC leptons, forming neutral AC atoms \cite{FKS}.
The second assumes a charge asymmetry  of the $O^{--}$ which forms the atom-like states with
primordial helium \cite{mpla,I}.

If new stable species belong to non-trivial representations of
the SU(2) electroweak group, sphaleron transitions at high temperatures
can provide the relation between baryon asymmetry and excess of
-2 charge stable species, as it was demonstrated in the case of WTC
\cite{KK,KK2,unesco,iwara}.

 After it is formed
in the Standard Big Bang Nucleosynthesis (BBN), $^4He$ screens the
$O^{--}$ charged particles in composite $(^4He^{++}O^{--})$ {\it
$OHe$} ``atoms'' \cite{I}.
In all the models of $OHe$, $O^{--}$ behaves either as a lepton or
as a specific ``heavy quark cluster" with strongly suppressed hadronic
interactions.
The cosmological scenario of the $OHe$ Universe involves only one parameter
of new physics $-$ the mass of O$^{--}$. Such a scenario is insensitive to the properties of $O^{--}$ (except for its mass), since the main features of the $OHe$ dark atoms are determined by their nuclear interacting helium shell. In terrestrial matter such dark matter species are slowed down and cannot cause significant nuclear recoil in the underground detectors, making them elusive in direct WIMP search experiments (where detection is based on nuclear recoil) such as CDMS, XENON100 and LUX. The positive results of DAMA experiments (see \cite{DAMAtalk} for review and references) can find in this scenario a nontrivial explanation due to a low energy radiative capture of $OHe$ by intermediate mass nuclei~\cite{mpla,DMRev,DDMRev}. This explains the negative results of the XENON100 and LUX experiments. The rate of this capture is
proportional to the temperature: this leads to a suppression of this effect in cryogenic
detectors, such as CDMS. 

OHe collisions in the central part of the Galaxy lead to OHe
excitations, and de-excitations with pair production in E0 transitions can explain the
excess of the positron-annihilation line, observed by INTEGRAL in the galactic bulge \cite{DMRev,DDMRev,KK2,CKWahe}.

One should note that the nuclear physics of OHe is in the course of development and its basic element for a successful and self-consistent OHe dark matter scenario is related to the existence of a dipole Coulomb barrier, arising in the process of OHe-nucleus interaction and providing the dominance of elastic collisions of OHe with nuclei. This problem is the main open question of composite dark matter, which implies correct quantum mechanical solution \cite{CKW}. The lack of such a barrier and essential contribution of inelastic OHe-nucleus processes seem to lead to inevitable overproduction of anomalous isotopes \cite{CKW2}.

Production of pairs of elementary stable doubly charged heavy leptons is characterized by a number of distinct experimental signatures that would provide effective search for
them at the experiments at the LHC and test the atom-like structure of the cosmological dark matter. Moreover, astrophysical consequences of composite dark matter model can reproduce experimentally detected excess in positron annihilation line and high energy positron fraction in cosmic rays only if the mass of double charged $X$ particles is in the 1 TeV range (Section 2). We discuss confrontation of these predictions and experimental data in Section 3. The current status and expected progress in experimental searches for stable double charged particles as constituents of composite dark matter are summarized in the concluding Section 4. 
\section{Indirect effects of composite dark matter}
\label{astro}
The existence of such form of matter as O-helium should lead to a number of astrophysical signatures,
which can constrain or prove this hypothesis. One of the signatures of O-helium can be a presence
of an anomalous low Z/A component in the cosmic ray flux. O-helium atoms that are present in the
Galaxy in the form of the dark matter can be destroyed in astrophysical processes and free $X^{−−}$ can be
accelerated as ordinary charged particles. O-helium atoms can be ionized due to nuclear interaction with
cosmic rays or in the front of a shock wave in the Supernova explosions, where they were effectively
accumulated during star evolution \cite{I}. If the mechanisms of $X^{−−}$ acceleration are effective, the
low Z/A component with charge −2 should be present in cosmic rays at the level of $F_X/F_p \sim 10^{-9}m^{-1}_o $ \cite{KK2},
and might be observed by PAMELA and AMS02 cosmic ray experiments. Here $m_o$ is the mass
of O-helium in TeV, $F_X$ and $F_p$ are the fluxes of $X^{−−}$ and protons, respectively.
\subsection{Excess of positron annihilation line in the galactic bulge}
\label{bulge}
\begin{figure}[htbp]
    \begin{center}
        \includegraphics[scale=1]{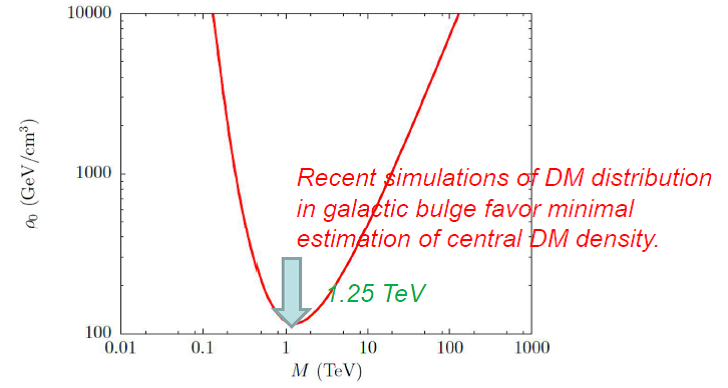}
        \caption{Dark matter is subdominant in the central part of Galaxy. It strongly suppresses it dynamical effect and causes large uncertainty in dark matter density and velocity distribution. At this uncertainty one can explain the positron line excess, observed by INTERGRAL, for a wide range of $m_o$ given by the curve with minimum at $m_o = 1.25 \TeV$. However, recent analysis of possible dark matter distribution in the galactic bulge favor minimal value of its central density.}
        \label{integral}
    \end{center}
\end{figure}

Another signature of O-helium in the Galaxy is the excess of the positron annihilation line in cosmic
gamma radiation due to de-excitation of the O-helium after its interaction in the interstellar space. If
2S level of O-helium is excited, its direct one-photon transition to the 1S ground state is forbidden and
the de-excitation mainly goes through direct pair production. In principle this mechanism of positron
production can explain the excess in positron annihilation line from the galactic bulge, measured by the
INTEGRAL experiment. Due to the large uncertainty of DM distribution in the galactic bulge this interpretation of the INTEGRAL data is possible in a wide range of masses of
O-helium with the minimal required central density of O-helium dark matter at $m_o = 1.25 \TeV$ \cite{integral1,integral2}
For the smaller values of $m_o$ on needs larger central density to provide effective excitation of O-helium in collisions
Current analysis favors lowest values of central dark matter density, making possible O-helium explanation for this excess only for a narrow window around this minimal value (see Fig. \ref{integral})

\subsection{Composite dark matter solution for high energy positron excess}
\label{HEpositrons}
\begin{figure}[htbp]
    \begin{center}
        \includegraphics[scale=1]{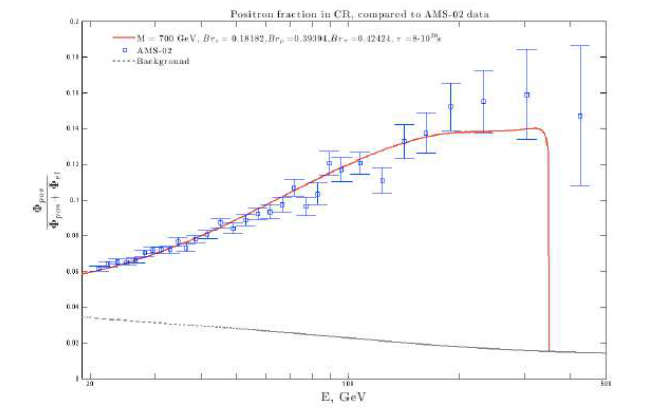}
        \caption{Best fit high energy positron fluxes from decaying composite dark matter in confrontation with the results of AMS02 experiment.}
        \label{ams}
    \end{center}
\end{figure}

In a two-component dark atom model, based on Walking Technicolor, a
sparse WIMP-like component of atom-like state, made of positive and neg-
ative doubly charged techniparticles, is present together with the dominant
OHe dark atom and the decays of doubly positive charged techniparticles to
pairs of same-sign leptons can explain the excess of high-energy cosmic-ray
positrons, found in PAMELA and AMS02 experiments [17]. This explana-
tion is possible for the mass of decaying +2 charged particle below 1 TeV
and depends on the branching ratios of leptonic channels (See Fig. \ref{ams}).

Since even pure
lepton decay channels are inevitably accompanied by gamma radiation the
important constraint on this model follows from the measurement of cosmic
gamma ray background in FERMI/LAT experiment. 
\begin{figure}[htbp]
    \begin{center}
        \includegraphics[scale=1.2]{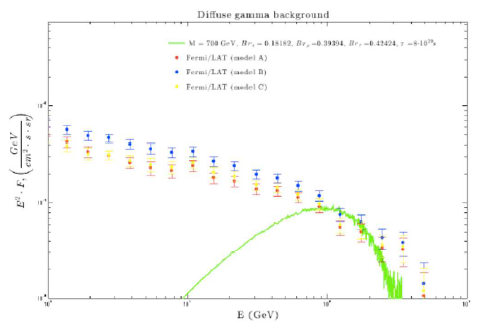}
        \caption{Gamma ray flux accompanying the best fit high energy positron fluxes from decaying composite dark matter reproducing the results of AMS02 experiment, in confrontation with FERMI/LAT measurement of gamma ray background.}
        \label{fermi}
    \end{center}
\end{figure}
The multi-parameter
analysis of decaying dark atom constituent model determines the maximal model independent value of the mass of decaying
+2 charge particle, at which this explanation is possible $$m_o<1 TeV.$$
One should take into account that according to even in this range hypothesis on decaying composite dark matter, distributed in the galactic halo, can lead to gamma ray flux exceeding the measured by FERMI/LAT.

\subsection{Sensitivity of indirect effect of composite dark matter to the mass of their double charged constituents}
\label{mass}
We see that indirect effects of composite dark matter strongly depend on the mass of its double charged constituents. 

To explain the excess of positron annihilation line in the galactic bulge mass of double charged constituent of O-helium should be in a narrow window around
$$m_o = 1.25 \TeV.$$

To explain the excess of high energy cosmic ray positrons by decays of constituents of composite dark matter with charge +2 and to avoid overproduction of gamma background, accompanying such decays, the mass of such constituent should be in the range
$$m_o < 1 \TeV.$$

These predictions should be confronted with the experimental data on the accelerator search for stable double charged particles.
\section{Searches for stable multi-charged particles}
\label{experiment}
A~new charged massive particle with electric charge $\neq 1e$ would represent a~dramatic deviation from the~predictions of the~Standard Model, and such a~spectacular discovery would lead to fundamental insights and critical theoretical developments. Searches for such kind of particles were carried out in many cosmic ray and collider experiments (see for instance review in~\cite{fair}). 
Experimental search for double charged particles is of a~special interest because of important cosmological sequences discussed in previous sections. In a~tree approximation, such particles cannot decay to pair of quarks due to electric charge conservation and only decays to the~same sign leptons are possible. The~latter implies lepton number nonconservation, being a~profound signature of new physics. In general, it makes such states sufficiently long-living in a~cosmological scale.  

Obviously, such experiments are limited to look only for the~``long-lived'' particles, which do not decay within a~detector, as opposed to truly stable particles, which do not decay at all. Since the~kinematics and cross sections for double charged particle production processes cannot be reliably predicted, search results at collider experiments are usually quoted as cross section limits for a~range of charges and masses for well-defined kinematic models. In these experiments, the~mass limit was set at the~level of up to $100$~\GeV(see e.g. for review~\cite{fair}).

The~CDF experiment collaboration performed a~search for long-lived double charged Higgs bosons ($H^{++}, H^{--}$) with $292$~pb$^{-1}$ of data collected in $p\bar{p}$ collisions at $\sqrt{s}=1.96$~\TeV{}~\cite{Acosta:2005np}. 
The~dominant production mode considered was $p\bar{p}\rightarrow \gamma^{\star}/Z+X\rightarrow H^{++}H^{--}+X$. 

Background sources include jets fragmenting to high-$\ensuremath{p_{\text{T}}}$ tracks, $Z\rightarrow ee$, $Z\rightarrow \mu \mu$, and $Z\rightarrow \tau \tau$, where at least one $\tau$ decays hadronically. Number of events expected from these backgrounds in the~signal region was estimated to be $<10^{-5}$ at $68\%$ confidence level (CL).

\begin{figure}[htbp]
    \begin{center}
        \includegraphics[scale=1.2]{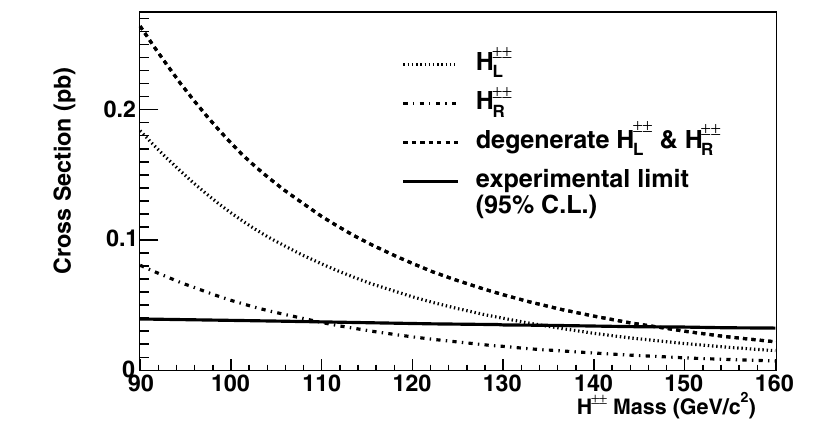}
        \caption{The~comparison of the~experimental cross section
upper limit with the~theoretical next-to-leading order cross
section for pair production of $H^{\pm\pm}$ bosons. The~theoretical
cross sections are computed separately for bosons with left-handed ($H^{\pm\pm}_L$) and right-handed ($H^{\pm\pm}_R$) couplings, and summed
for the~case that their masses are degenerate,~\cite{Acosta:2005np}.}
        \label{CDF_limits}
    \end{center}
\end{figure}

Not a~single event with a~$H^{++}$ or $H^{--}$ was found in experimental data. This allows to set cross-section limit shown in Fig.~\ref{CDF_limits} as a~horizontal solid line.
Next-to-leading order theoretical calculations of the cross-section for pair production of $H^{\pm\pm}$ bosons for left-handed and right-handed couplings are also shown in this figure. 
Comparison of expected and observed cross-section limits gives the~following mass constrains: $133$ and $109$~\GeV{} on the~masses of long-lived $H^{\pm\pm}_L$ and $H^{\pm\pm}_R$, respectively, at $95\%$ CL as shown in Fig.~\ref{CDF_limits}. 
In case of degenerate $H^{\pm\pm}_L$ and $H^{\pm\pm}_R$ bosons, the~mass limit was set to $146$~\GeV{}. 

\subsection{Searches at Large Hadron Collider}
Significant increase of beam energy at the~Large Hadron Collider (LHC) opens a~new era in the~high energy physics and allows accessing uncharted territories of particle masses. In this section the~results of searches for the~multi-charged particles, performed by the~ATLAS and the~CMS collaborations at LHC, will be described. 

Calculations of the cross sections assume that these particles are generated as new massive spin-$1/2$ ones, neutral under SU(3)$_C$ and SU(2)$_L$.

\subsubsection{ATLAS experiment at LHC}
\label{ATLAS_search}
In Run~1 (2010--2012), the~ATLAS~\cite{Aad:2008zzm} collaboration at LHC performed two searches for long-lived multi-charged particles, including the~double charged particles: 
one search with $4.4$~fb$^{-1}$ of data collected in $pp$ collisions at $\sqrt{s}=7$~\TeV{}~\cite{Aad:2013pqd}, 
and another one with $20.3$~fb$^{-1}$ collected at $\sqrt{s}=8$~\TeV{}~\cite{Aad:2015oga}. 

Both these searches feature particles with large transverse momentum values, traversing the~entire ATLAS detector. An~energy loss of a~double charged particle is by a~factor of $q^2=4$ higher than that of single charged particle. Such particles will leave a~very characteristic signature of high ionization in the~detector. 
More specifically, the~searches look for particles with anomalously high ionization on their tracks in three independent detector subsystems: silicon pixel detector (Pixel) and transition radiation tracker (TRT) in the~ATLAS inner detector, and monitoring drift tubes (MDT) in the~muon system.

The~estimate of the~expected background originating from the~SM processes and providing input into the~signal region D was calculated to be $0.013\pm0.002 \textrm{(stat.)}\pm0.003\textrm{(syst.)}$ events. 

In order to define cross section, a~reconstruction efficiency of signal particles has to be known.
\begin{figure}[htbp]
    \begin{center}
        \includegraphics[scale=0.45]{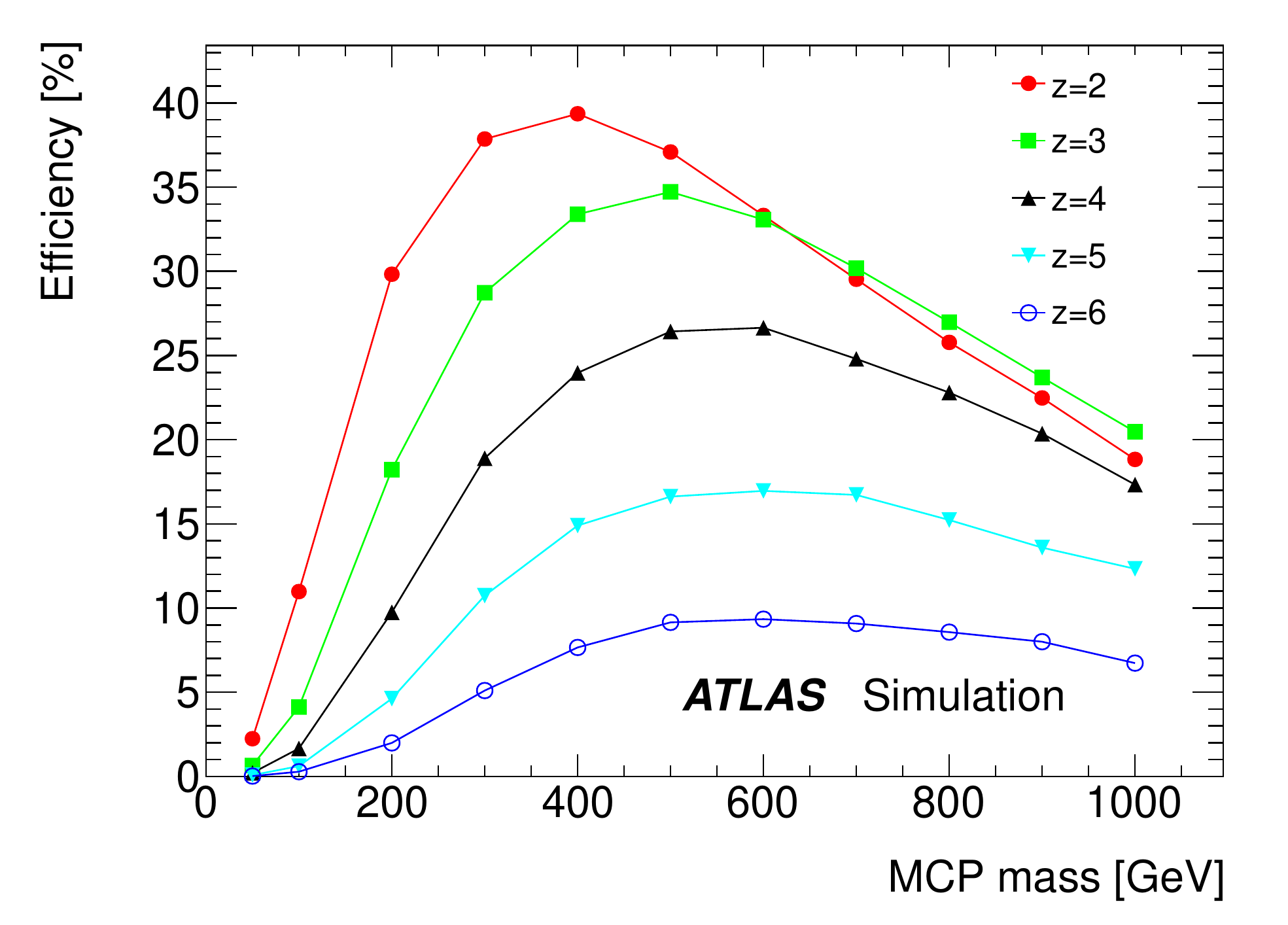}
        \caption{The~signal efficiencies for different masses and charges of the~multi-charged particles for the~DY production model. Double charged particles are denoted as ``$z=2$'' (red points and line). The~picture is taken from~\cite{Aad:2015oga}.}
        \label{ATLAS_EfficiencyTrend}
    \end{center}
\end{figure}
This value is defined as a~fraction of simulated events with at least one multi-charged particle satisfying all of the~aforementioned criteria over the~number of all generated events. In other words, it is a~search sensitivity to find a~hypothetical particle with the~ATLAS experiment. These values are shown in Fig.~\ref{ATLAS_EfficiencyTrend} for each considered signal sample containing particles with charges from $2$ to $6$. 

\begin{figure}[htbp]
    \begin{center}
        \includegraphics[scale=0.45]{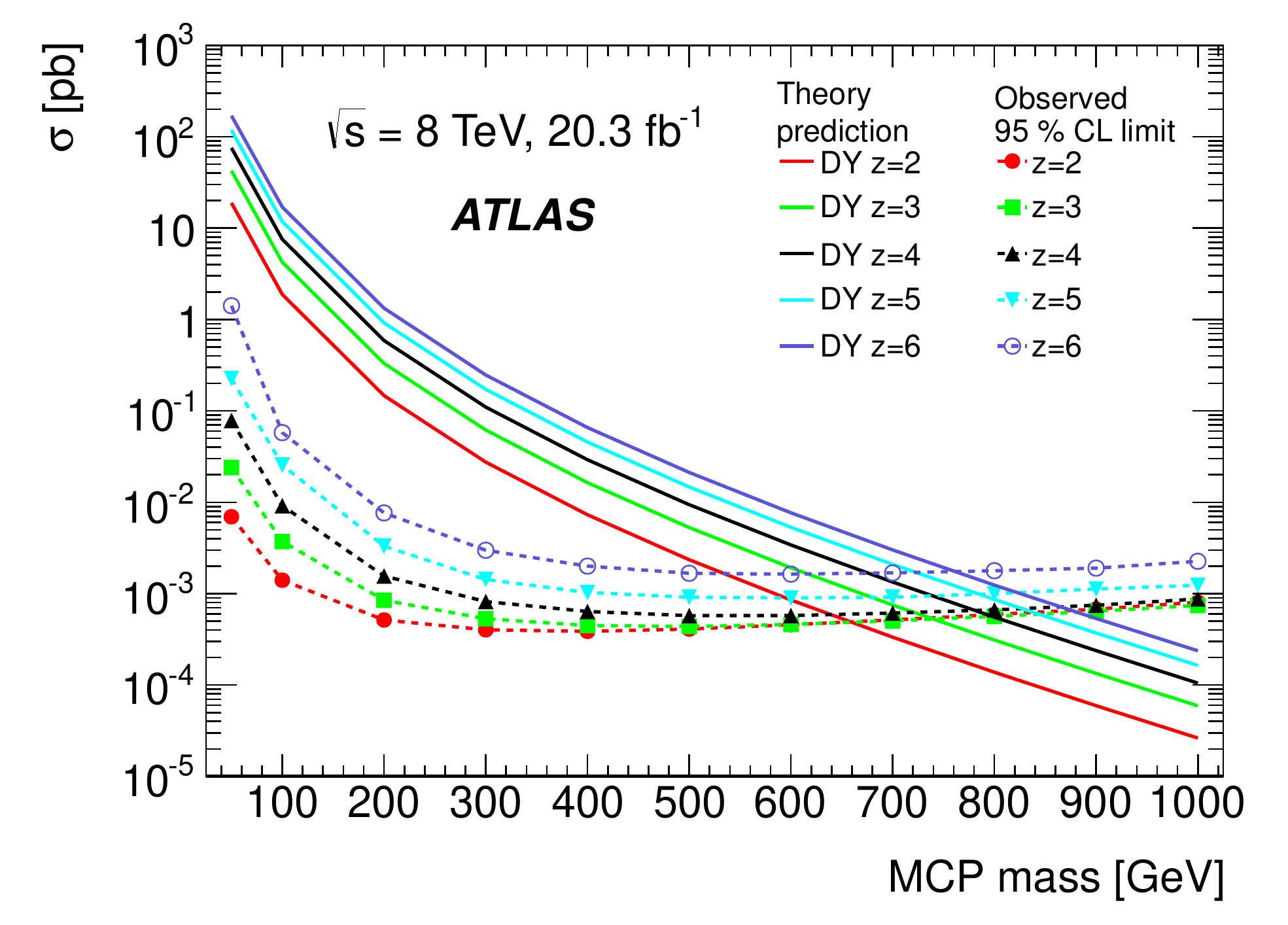}
        \caption{Observed $95\%$ CL cross-section upper limits and theoretical cross-sections as functions of the multi-charged particles mass. Again, the~double charged particles are denoted as ``$z=2$'' (red points and lines). The~picture is taken from~\cite{Aad:2015oga}.}
        \label{ATLAS_Limits}
    \end{center}
\end{figure}
No events with double charged particles were found in neither 2011 or 2012 experimental data sets, setting the~lower mass limits to $430$ and $660$~\GeV{}, respectively, at $95\%$ CL. The~comparison of observed cross-section upper limits and theoretically predicted cross-sections is shown in Fig.~\ref{ATLAS_Limits}.
\subsubsection {CMS experiment at LHC}
\label{CMS_search}
In parallel to the~ATLAS experiment, the~CMS~\cite{Chatrchyan:2008aa} collaboration at LHC performed a~search for double charged particles, using  $5.0$~fb$^{-1}$ of data collected in $pp$ collisions at $\sqrt{s}=7$~\TeV{} and $18.8$~fb$^{-1}$ collected at $\sqrt{s}=8$~\TeV{}~\cite{Chatrchyan:2013oca}.

The~search features particles with high ionization along their tracks in the~inner silicon pixel and strip tracker. Tracks with specific ionization $I_h>3$~\MeV{}/cm were selected. The muon system was used to measure the~time-of-flight values. Tracks with $1/\beta>1.2$ were considered.

For the~part of the~search based on the~$\sqrt{s}=7$~\TeV{} data, the~number of events in the~signal region, expected from SM processes, was estimated to be $0.15\pm0.04$, whereas for the~$\sqrt{s}=8$~\TeV{} part it was $0.52\pm0.11$~events. 
The~uncertainties include both statistical and systematical contributions. $0$ and $1$~events were observed in the~signal regions for the~$7$ and $8$~\TeV{} analyses, respectively, which is consistent with the~predicted event rate.

\begin{figure}[htbp]
    \begin{center}
        \includegraphics[scale=0.6]{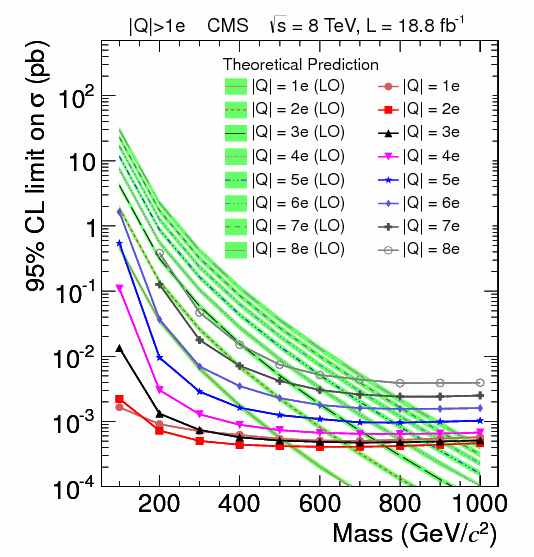}
        \caption{Observed $95\%$ CL cross-section upper limits and theoretical cross-sections as functions of the multi-charged particles mass in CMS search at the~$\sqrt{s}=8$~\TeV{}. The~double charged particles are denoted as ``$|Q|=2e$''. The~picture is taken from~\cite{Chatrchyan:2013oca}.}
        \label{CMS_limits_8TeV}
    \end{center}
\end{figure}

Comparison between observed upper cross section limits and theoretically predicted cross section values for the~$8$~\TeV{} is shown in Fig.~\ref{CMS_limits_8TeV}.

For the~$8$~\TeV{} search, the~mass limit of $665$~\GeV{} was obtained. This result (within uncertainties) is very close to the~ATLAS limit of $660$~\GeV{} for the~$8$~\TeV{}  data set. Also, CMS treated the~results obtained at $7$ and $8$~\TeV{} as combined. This allowed to push the~lower mass limit to $685$~\GeV{} at $95\%$ CL. A~combination of the~results of two experiments for $8$~\TeV{} would  mean an~increase of statistics by a~factor of $2$. Having said that, one can conclude that the~mass limit based on the~results of both experiment for double charged particles can be set at the~level of about $730$~\GeV{}.

\subsubsection {What one expects from LHC Run~2}
\label{LHC Run 2}

Considering a~recent CMS Physics Analysis Summary~\cite{CMS:2016ybj} and an~ATLAS paper in preparation, both on the~searches for double charged particles in data delivered by LHC to these experiments in 2015--2016, we may conclude that each of these two experiments will be able to set a~lower mass limit on the~double charged particles at $m=1000\pm50$~\GeV{}. 
Going further and considering the data taking periods of ATLAS and CMS until the end of Run 2 (end of 2018), we can estimate a~low limit on the~double charged particles mass corresponding to the~Run 2 data set. Several assumptions are made in these speculations:
\begin{itemize}
	\item by the~end of 2018, ATLAS and CMS will each record about $120$~fb$^{-1}$ of $\sqrt{s}=13$~\TeV{} data;
    \item signal efficiency will remain at a~present level in both experiments, without being compromised by high detector occupancy or any other effects;
    \item no double charged particle candidates will be in observed in the~first place.
\end{itemize}

Considering all of the~above, the~ATLAS and CMS collaborations may each be expected to set the~lower mass limits at the~level of $1.2$~\TeV{} based on their analyses of the~entire $13$~\TeV{} data set. If these two experiments combined their independently gathered statistics together for this kind of search, the~limits would go as high as up to $1.3$~\TeV{}.
\section{Conclusions}
The existence of heavy stable neutral particles is one of the popular solutions for the dark matter problem. However, DM considered to be electrically neutral, potentially  can be formed by stable heavy charged particles bound in neutral atom-like states by Coulomb attraction. Analysis of the cosmological data and atomic composition of the Universe gives the constrains on the particle charge
showing that only -2 charged constituents, being trapped by primordial helium in neutral O-helium states, can avoid the problem of overproduction of the anomalous isotopes of chemical elements, which
are severely constrained by observations. Cosmological model of O-helium dark matter can even explain
puzzles of direct dark matter searches.

Stable  charge -2 states ($X^{--}$) can be elementary like AC-leptons or technileptons, or look like
technibaryons. The latter, composed of techniquarks, reveal their structure at much higher energy scale
and should be produced at colliders and accelerators as elementary species. They can also be composite like ``heavy quark clusters'' $\bar U \bar U \bar U$ formed by anti-U quark in one of the models of fourth generation \cite{I} or $\bar u_5 \bar u_5 \bar u_5$ of
(anti)quarks $\bar u_5$ of stable 5th family in the approach \cite{Norma}.

In the context of composite dark matter scenario accelerator search for stable doubly charged leptons
acquires the meaning of direct critical test for existence of charged constituents of cosmological dark matter.

The signature for AC leptons and techniparticles is unique and distinctive what allows to separate them from other hypothetical exotic particles. Composite dark matter models can explain observed excess of high energy positrons and gamma radiation in positron annihilation line at the masses of $X^{--}$ in the range of $1$~\TeV{}, it makes search for double charged particles in this range direct experimental test for these predictions of composite dark matter models.

Test for composite $X^{--}$ can be only indirect: through the search for heavy hadrons, composed of single $U$ or $\bar U$ and light quarks (similar to R-hadrons).
 
The~ATLAS and CMS collaborations at the~Large Hadron Collider are searching for the~double charged particles since 2011. The~most stringent results achieved so far exclude the~existence of such particles up to their mass of $680$~\GeV{}. This value was obtained by both ATLAS and CMS collaborations independently. It is expected that if these two collaborations combine their independently gathered statistics of LHC Run 2 (2015--2018), the~lower mass limit of double charged particles could reach the~level of about $1.3$~\TeV{}.
\section*{Acknowledgements}
We thank K.Belotsky, J.-R. Cudell, C. Kouvaris, M.Laletin for collaboration in development of the presented approach. The work was supported by Russian Science Foundation and fulfilled in the framework of MEPhI
Academic Excellence Project (contract 02.a03.21.0005, 27.08.2013)..


\end{document}